\documentclass{aastex63}
\usepackage{lineno}

\newcommand\tempo{T\textsc{empo}2}     
\newcommand\toa{TOA}

\usepackage[normalem]{ulem}

\received{June 1, 2019}
\revised{January 10, 2019}
\accepted{\today}
\submitjournal{AJ}

\shorttitle{Open Source PEP}
\shortauthors{Chandler et al.}
\graphicspath{{./}{figures/}}

\begin{document}

\title{The Planetary Ephemeris Program: Capability, Comparison, and Open Source Availability}

\correspondingauthor{Robert D. Reasenberg}
\email{rreasenberg@ucsd.edu}

\nocollaboration{6}

\author{John F. Chandler}
\affiliation{University of California, San Diego, 9500 Gilman Dr, La Jolla, CA 92093, USA\\}
\affiliation{Center for Astrophysics, Harvard and Smithsonian, 60 Garden St, Cambridge, MA 02138, USA\\}

\author{James B. R. Battat}
\affiliation{Dept. of Physics, Wellesley College, 106 Central Street, Wellesley, MA 02481, USA\\}

\author{Thomas W. Murphy, Jr.}
\affiliation{University of California, San Diego, 9500 Gilman Dr, La Jolla, CA 92093, USA\\}

\author{Daniel Reardon}
\affiliation{Centre for Astrophysics and Supercomputing, Swinburne Technical University, John Street, Hawthorn Victoria 3122, Australia\\
}

\author{Robert D. Reasenberg}
\affiliation{University of California, San Diego, 9500 Gilman Dr, La Jolla, CA 92093, USA\\}
\affiliation{Center for Astrophysics, Harvard and Smithsonian, 60 Garden St, Cambridge, MA 02138, USA\\}

\author{Irwin I. Shapiro}
\affiliation{Center for Astrophysics, Harvard and Smithsonian, 60 Garden St, Cambridge, MA 02138, USA\\}

\begin{abstract}
We describe for the first time in the scientific literature the Planetary Ephemeris Program (PEP), an open-source general-purpose astrometric data analysis program. We discuss, in particular, the implementation of pulsar timing analysis, which was recently upgraded in PEP to handle more options. This implementation  was done independently of other pulsar  programs, with minor exceptions that we discuss. We illustrate the implementation of this capability by comparing the post-fit residuals from the analyses of time-of-arrival observations by both PEP and \tempo{}. The comparison shows substantial agreement: 22\,ns rms differences for 1,065 pulse time-of-arrival measurements for the millisecond pulsar in a binary system, PSR J1909-3744 (pulse period 2.947108\,ms; full-width half-maximum of pulse 43\,$\mu s$) for epochs in the interval from December 2002 to February 2011.

\end{abstract}


\section{Introduction} \label{sec:intro}

This paper provides the first published description of the Planetary Ephemeris Program (PEP), which has been in use and evolving for nearly six decades. PEP is a general-purpose astrometric data-analysis program that can compute numerical ephemerides and can analyze simultaneously a heterogeneous collection of astrometric data. Very few programs like PEP exist, and, to our knowledge, PEP is the only one that has been freely available, including source code, to interested parties. Among the other programs with comparable scope are DPODP (JPL, USA) \citep{Park_2021}, INPOP (IMCCE Obs. Paris, France) \citep{Viswanathan_2018}, the analysis model at the Institut f{\"u}r Erdmessung (IfE), Leibniz Universit{\"a}t Hannover, Germany \citep{2019JGeod..93.2195M}, and EPM (Laboratory of Ephemeris Astronomy, Russia) \citep{2014CeMDA.119..237P}.

Ephemeris and modeling codes such as PEP are flexible tools that have played a critical role in determining physical parameters of solar-system bodies such as orbital initial conditions, planetary masses, and the length scale (the AU) in terrestrial length units \citep{AJ.72.338}, and in testing fundamental physics theories. PEP has a particularly rich history with the latter. It has been used to measure general relativistic effects such as the de Sitter precession of the moon \citep{PhysRevLett.61.2643} and the Shapiro time delay \citep{ApJ.234.L219}. In addition, PEP has been used to test physics theories that go ``beyond the standard model''.

Over the past fifty years, the suite of solar-system astrometric data sets has grown in scope and precision, and advancements from the theory community have triggered an evolution in the types of scientific questions that we ask of the data. In response to these changes, PEP has been updated, both in terms of the types of observables it can analyze, and in terms of the precision with which it can analyze observables. A major current effort with PEP is the analysis of the Lunar Laser Ranging (LLR) data from the Apache Point Observatory Lunar Laser-ranging Operation experiment (APOLLO) \citep{Murphy:2007ed, Murphy:2012rea} in conjunction with the other sets of data described in Table~\ref{tab:Data}. APOLLO represents an order-of-magnitude improvement in the precision of LLR, yielding an Earth-Moon distance equivalent of 1.2\,mm median uncertainty in a single one hour observing session \citep{2009PASP..121...29B, Liang:2017xjb}.  As is frequently true of the analysis of new (improved) astrometric data, the model of the motion and rotation of the Moon needed to reduce these data to near their noise level likely has more free parameters than can be estimated simultaneously. This is in general true of the solar-system model and data set. Under these conditions, it may be useful to add auxiliary data types that likely will reduce the degeneracy of the estimator, and thus allow for the use of a more complex model.  Therefore, driven by this advance in LLR, PEP has been upgraded to handle pulse time-of-arrival (\toa{}) data from more types of pulsars than before, such as those with massive companions and those with complicated dispersion histories.

Such an upgrade benefits other analyses, too, and opens the possibility of new ones.  For example, pulsar \toa{} data can be used in conjunction with VLBI data to tie PEP's planetary reference frame to the extra-galactic radio reference frame \citep{1996AJ....112.1690B}.  Moreover, even without a frame tie, the observation of pulsars well away from the ecliptic offers a breaking of the degeneracy implicit in relying on data restricted to the vicinity of a single plane.  For another example, the Dvali-Gabadadze-Porrati (DGP) model of gravity \citep{Dvali:2000hr}, which was introduced to explain cosmic acceleration without dark energy, predicts a common-mode precession of all orbiting bodies in the solar system by a fixed angular rate. For co-planar orbits, this looks like a net rotation. The inclinations of the planetary orbits relative to the ecliptic break this degeneracy, but the small inclination angles strongly suppress the sensitivity to the predicted effect in DGP theory \citep{Battat:2008bu}. The out-of-plane perspective provided by pulsar timing would enhance the sensitivity to DGP-like anomalies.

\begin{deluxetable*}{lllll}       
\tablecaption{Astrometric Data Currently in Use with PEP \label{tab:Data}}
\tablewidth{0pt}
\tablehead{
\colhead{Data sets} & \colhead{Time range} & \colhead{Number}  &  \colhead{Uncertainties} & \colhead{Units}}
\startdata
APOLLO NPs\tablenotemark{c}            &   2006-2020  & 3397    & 6 to 470 & ps\\
Other LLR NPs           &  1969-2003  & 14400    &  0.04 to 900 & ns\\
Mercury radar ranges    &  1969-1997  & 8054     & 0.1 to 40 & $\mu$s\\
MESSENGER NPs           &  2011-2014  & 1311     & 40 & ns\\
Venus radar ranges      &  1969-1982  & 5674     & 0.1 to 20 & $\mu$s\\
Mariner 9 NPs           &  1971-1972  &  185     & 0.2 to 10 & $\mu$s\\
Viking orbiter ranges (S and X bands)& 1976-1980  & 1041     & 14 to 160 & ns\\
Viking lander ranges (X band only)   &  1980-1982  &  239     & 35 to 410 & ns\\
Mars Pathfinder ranges  &  1997-1997   &  90     & 67 to 150 & ns\\
Mars Global Surveyor NPs&  1999-2006   & 164562  & 40 to  40 & ns\\
Mars Odyssey NPs        &  2002-2008   & 293594  & 20 to  20 & ns\\
Outer planet NPs        &  1973-1981   &   6     & 27 to 420 & $\mu$s\\
\enddata
\tablenotetext{c}{Normal Points, see Section~\ref{sec:Grow}.}
\end{deluxetable*}

In Section~\ref{sec:PEP}, we provide an overview of PEP, including a description of the astrometric data sets it can process. This section concludes with a discussion of PEP's availability, both historically and now plus the foreseeable future. In Section~\ref{sec:Grow}, we discuss the handling of pulsar timing observables in PEP, and present the results of a  validation of this capability against \tempo{} \citep{Hobbs:2006cd,Edwards:2006zg}, a program specialized for the analysis of \toa{} observations of pulses transmitted from a pulsar and in wide use by the pulsar community.  Such comparisons of scientific codes are an essential part of the effort to ensure the validity of data analysis in an era with very large precision codes and evolving demands on these codes. A brief summary of the key points of the paper and some further observations are contained in Section~\ref{sec:Conc}, Conclusion. Finally, in the Appendix, we provide the algorithm that allows us to make numerous numerical solution estimates at the end of a data analysis (when the estimate is converged and most nearly linear).  Unlike most of the algorithms embedded in PEP, we believe this one does not have a counterpart in other analysis packages.

\begin{deluxetable*}{ll}
\tablecaption{Observables Supported by the Planetary Ephemeris Program\label{tab:Obs}}
\tablehead{
\colhead{Observable}  & 		\colhead{Highest precision data analyzed to date}
}
\startdata
Classical optical observations of stars and planets & 	(These data are no longer considered useful)\\
\hspace{6 mm}Transit circle	&			0.5 as\\
\hspace{6 mm}Photographic	&				1 as\\
\hspace{6 mm}Differential	&				20 mas\\
\hspace{6 mm}Occultation timing	&			0.3 s\\
Radar  &  \\
\hspace{6 mm}Lunar	&					N/A\\
\hspace{6 mm}Planetary	&				90 ns\\
Laser ranging\\
\hspace{6 mm}Artificial satellite	&			N/A\\
\hspace{6 mm}Lunar				&		6.5 ps  \hspace{4 mm}  (about 1 mm, one way)\tablenotemark{a}\\
Spacecraft  ranging\\
\hspace{6 mm}Interplanetary		&			60 ns   \\
\hspace{6 mm}Planetary orbiter	&			20 ns     \\      
\hspace{6 mm}Planetary lander	&			14 ns  \\
Very long baseline interferometry (VLBI, astrometric)& 	$<$ 1 mas\\
Pulsar timing			&			20 ns  \\
\enddata
\tablenotetext{a}{By convention, time is taken to be round-trip time (RTT) and distance as RTT c/2, where c is the speed of light. }
\end{deluxetable*}

\begin{deluxetable*}{ll}       
\tablecaption{Organization of the Planetary Ephemeris Program \label{tab:Org}}
\tablewidth{0pt}
\tablehead{
\colhead{Major Sections of PEP} & 		\colhead{Key Characteristics}\\
}
\startdata
Input	&	Initial conditions, initial values of model\\      
         &  \hspace{2 mm}parameters, data\\
Evaluation of analytic theories of body motion  &  Used when the complexity available from\\  & \hspace{2 mm}numerical integration is not required\\
Numerical integration		&		Fixed or variable step size, as needed\\
& \hspace{2 mm}forward and backward\\
& \hspace{2 mm}automatic starting procedure\\
\hspace{6 mm}Single Bodies\\
\hspace{6 mm}Lunar motion and rotation together\\
\hspace{6 mm}All major bodies of the solar system together, \\
\hspace{8 mm}including the moon\\
Calculation of Theoretical Observable   	&		Based on models and integrated ephemerides\\
\hspace{6 mm} Pre-fit residuals (O-C)\\
Parameter estimation		&	Iterated, linearized, weighted least squares (WLS)\\
\hspace{6 mm}Forming and Saving normal equations from a series of data\\
\hspace{6 mm}Restoring and combining normal equations series \\
\hspace{6 mm}Introduction of a priori constraints (simple, vector)\\
\hspace{6 mm}Solving normal equations  &  Matrix inversion and multiply\\
\hspace{10 mm}Partial Pre-Reduction\tablenotemark{b} \\
\hspace{10 mm}Iterative improvement of solutions\\
\hspace{6 mm}Post-fit residuals\\
\enddata
\tablenotetext{b}{Partial Pre-Reduction, see Appendix }
\end{deluxetable*}

\section{The Planetary Ephemeris Program} \label{sec:PEP}

PEP was originally created at MIT Lincoln Laboratory starting in the early 1960s. It was brought to the main campus half a decade later, which then evolved to be the center of PEP activity. In the early 1980s that center moved to the (then named) Harvard-Smithsonian Center for Astrophysics (CfA).  PEP is a general-purpose astrometric data-analysis program and can analyze simultaneously various kinds of astrometric data, as described in Table~\ref{tab:Obs}.  That table includes the highest precision of the data analyzed for each category, which sets the level of model complexity (completeness) required. A top-level outline of PEP is provided in Table~\ref{tab:Org}.

As might be expected from its inception era, PEP is written in Fortran, albeit modernized to make use of language features not available to the original Fortran II implementation. In the current era, we use the GNU G Fortran compiler.  Over the decades, PEP has accreted new capabilities as needed for its planned uses.  In Section~\ref{sec:Grow}, we discuss one such addition, the inclusion of the analysis of the time-of-arrival observations of pulses transmitted from a pulsar.  Although this addition uses algorithms generated within the PEP group, we are verifying that these algorithms are consistent with the best models used in the field.  To that end, we have compared the PEP analysis of a particular set of pulsar data with the corresponding analysis made with \tempo{}.

Over the past half century or so, PEP has been used in over 20 doctoral theses, including for example several on the fourth test of general relativity in the late 1960s and 1970s and in about 90 scientific papers including about 30 on tests of general relativity, competing theories and related cosmological issues such as a possible secular variation of the Newtonian ``constant" of gravitation. These range from tests of the Shapiro time delay during the early years \citep{PhysRevLett.20.1265, PhysRevLett.26.1132} to more recent papers on constraints on the Dvali-Gabadadze-Porrati braneworld theory of gravity \citep{PhysRevD.78.022003} and constraints on Standard-Model Extension Parameters \citep{PhysRevLett.99.241103}.

The maintainers of PEP have always made the PEP code available to members of the scientific community, thus starting well before the popular use of the term "open source."  Those interested generally had an affiliation with the then current  users and were directed to a member of the PEP community who would provide a copy of the code. We have since shifted to a more accessible and user-driven form of distribution. The PEP source code, utility programs, documentation, and unit tests are now all publicly available via GitLab\footnote{https://gitlab.com/jbattat/pep\_core}, and are distributed under a Creative  Commons Attribution-NonCommercial-ShareAlike license, making PEP open source in the modern sense. There are many technical memoranda written at MIT and CfA that describe algorithms and other key aspects of PEP. Several of these are available via GitLab, and the digital collection continues to grow.\footnote{https://gitlab.com/jbattat/pep\_doc}
Historically, PEP ran under various versions of IBM System 360 and System 370, then migrated to Sun OS, which became Solaris,
then jumped to Linux and thus to basically any type of Unix, including macOS.  It has been run under Windows experimentally, but not for production, as far as we know. \\

\section{Handling of pulsar \toa{} observations} \label{sec:Grow}

As for the pulsar \toa{} measurements, we also use Normal Points (NPs), each derived from observation of a pulsar, usually over the order of an hour or less (see below), rather than the raw data.  In creating these NPs, the received signal from the pulsar is recorded at a variety of frequencies.  In the Parkes Pulsar Timing Array Program, observations are conducted in the 20\,cm (1400\,MHz) band, or in the 10\,cm (3100\,MHz) and 40\,cm (700\,MHz) bands simultaneously \citep{Manchester:2012za}.  Often, the signal is recorded simultaneously in two bands, either widely spaced (e.g., two selected from these three major bands) or closely spaced (e.g., two narrow bands both within the 20\,cm band and separated only by about 60\,MHz to avoid a band of interference.  The number of observing sessions, each from about 10 to 60 minutes long, may range from one to ten in a day and the typical observing cadence is two weeks.  The individual pulse \toa{}s are then combined via integrating over time and frequency within each band during an observing session.  The stability of the pulsar allows such integrations to be made easily.  These combined (averaged) \toa{}s are the NPs, each representing the best estimate of the arrival time of a single pulse near the middle of an observing session and the middle of the frequency band over which the integration extended.

We have recently upgraded the PEP code for the \toa{} of the radio (or, in principle, other electromagnetic emission) pulses from a pulsar. The formulation of this observable has been published for the pulsar analysis packages \tempo{} and, more recently, PINT \citep{luo2021pint}. However, for our demonstration analysis, we chose to make only minor modifications to our decades-old simpler model that would work well enough for this purpose.  (The model accuracy needed to support the analysis of APOLLO NPs would not, for example, be sufficient to support the search for gravitational radiation in pulsar \toa{} data. That work is better done in a specialized package like PINT which, unlike \tempo{}, is designed to detect graviational waves. PINT is used by the NANOGrav Collaboration and by NASA's NICER Mission.)

We compared PEP to \tempo{} for real data, in particular the NPs from PSR J1909-3744, which span our entire c. 8-year interval of the observations.\footnote{Some of us are, separately, engaged in such a comparison between our PEP code and a similar, but less encompassing code - restricted to the solar system - developed at the Laboratory of Ephemeris Astronomy of the Institute of Applied Astronomy in Russia under the direction of Elena Pitjeva. The actual comparison is being carried out by Dmitry Pavlov from Russia and John Chandler from the U.S.} There was no consultation or coordination between the PEP and \tempo{} teams on algorithms used, except for the plasma, the reduction of station timing to UTC, and the treatment of the pulsar companion.  PSR J1909-3744, which is part of the binary system, was discovered in late 2001 \citep{Jacoby:2003nq} at the 64-meter-diameter Parkes radio telescope in Australia (see \cite{Manchester:2012za}).  The inter-pulse period is 2.95 ms.  The pulse itself has a full-width at half-maximum of 43\,$\mu$s, with a sharp peak, and is extremely stable.  These properties make this pulsar an excellent candidate for a software comparison.  

\subsection{Analysis of Data} \label{sec:Analysis}

First, we note that the 1,065 NPs used in both analyses were all based on data obtained at the Parkes Observatory.  The data analysis by the \tempo{} group was carried out using that program \citep{Edwards:2006zg}, and made use of JPL’s solar-system ephemeris, DE421 \citep{DE421}.  In Table~\ref{tab:results}, we show the results from this group’s analysis for 10 parameters of the pulsar system, with each parameter being described in the table.

The analysis with PEP  used the same DE421 ephemeris, and the same other relevant model parameters, such as for clock behavior earth-orientation and telescope coordinates. This group obtained corresponding parameter values, which are also shown in Table~\ref{tab:results}.  Finally, in addition, the parameter-estimate differences are given in this table.  All of these differences, except one, are well within the uncertainties.  

The computer programs of both groups use an iterated weighted-least-squares estimator in their analyses. 

We also note that there were many program changes needed in PEP to carry out this comparison.  For example, PEP’s model for interstellar dispersion was enhanced by the addition of an adjustable, constant time rate of change, and handling of pulsar companions was altered so that the effect could be either computed in PEP or imported from an external source; in the case of the comparison reported here it was imported from \tempo{}.

\begin{deluxetable*}{lrrrr}
\tablenum{4}
	\tablecaption{Estimated Parameter Values \label{tab:results}}
	\tablewidth{0pt}
	\tablehead{
	\colhead{Parameter} & \colhead{PEP} & \colhead{\tempo{}} & \colhead{Difference\tablenotemark{d}} & \colhead{Units}
	}
	\startdata
		Pulsar RA &	$(287447644473 \pm 1.3) \times 10^{-9}$ &	$(287447644474 \pm 1.3)\times 10^{-9}$ &	$-1\times 10^{-9}$ &	deg\\
        Pulsar Dec & $(-37737351861 \pm 4)\times 10^{-9}$ &	$(-37737351864 \pm 4) \times 10^{-9}$ &	$3\times 10^{-9}$ &	deg\\
        Parallax &	$0.8848 \pm 0.014$ &	$0.8864 \pm 0.014$ &	$-0.0026$ &	mas\\
        RA Proper Motion &	$-9.5339 \pm 0.0014$ & $-9.5332 \pm 0.0014$ & $-0.0007$ &	mas/yr\\
        Dec Proper Motion &	-35.7414 ± 0.0053 &	-35.7426 ± 0.0053 &	0.0012 & mas/yr\\
        Dispersion &	$(10393053 \pm 9)\times 10^{-6}$ & $(10393054 \pm 9)\times 10^{-6}$ &	$1\times 10^{-6}$ &	(elec/cc)pc\\
        Dispersion Rate & $(-331.1 \pm 5.1)\times 10^{-6}$ & $(-332.6 \pm 5.2)\times 10^{-6}$ & $1.5\times 10^{-6}$ & (elec/cc)pc/yr\\
        Pulse Offset & $-20  \pm  20$  &  $-20  \pm  21$  &  $0$  &  ns  \\
        Pulse Period & $(29471080234652320 \pm 9.4)\times 10^{-16}$ &	$(29471080234652280 \pm 9.4)\times 10^{-16}$ &	$40\times 10^{-16}$ &	msec\\
        Pulse Period Rate &	$(14025476 \pm 7.5)\times 10^{-27}$ & $(14025478 \pm 7.6)\times 10^{-27}$ & $-2\times 10^{-27}$ &	msec/sec\\
		\enddata
	\tablenotetext{d}{Value(PEP)-Value(\tempo{})}
\end{deluxetable*}

\subsection{Comparison of Analyses} \label{sec:Comp}

As can be seen in Table~\ref{tab:results}, the two programs produce nearly the same formal standard deviations for all 10 parameters to the number of places shown.  In three cases (Dispersion Rate, Pulse Offset, Pulse Period Rate), they differ by one in the least significant digit, which we do not take to be cause for concern.  Similarly, the two programs produce nearly the same parameter estimates. In all but one case, the absolute difference divided by the standard deviation is under or well under 1.  The exception is the Pulse Period which is discussed below.

It is worth noting that the parameter estimates are all sensitive to
context.  For example, the adoption of a particular planetary
ephemeris establishes the reference frame in which the position,
proper motion, and parallax of the pulsar are measured.  Accordingly,
it was essential for us to use the same ephemeris in order to
obtain comparable results.  Needless to say, it was also obligatory to
adopt matching reference epochs for the time-varying quantities
(position, dispersion, and pulse phase).  Indeed, we used a
single reference epoch for all three.  One remaining complication is
the time scale.  \tempo{} calculations are expressed in terms of the TCB
(Barycentric Coordinate Time) scale, while PEP and the DE421 ephemeris
use TDB (Barycentric Dynamical Time), and the two scales progress at
different rates.  The pulse period is the only parameter among the ten
we compared which is directly affected by this complication.  To make
the comparison in Table~\ref{tab:results}, we converted the \tempo{} result for the
pulse frequency expressed in pulses per TCB second into a pulse period
in TDB seconds, using a scale factor $(L_B)$ extracted from \tempo{}, but
there are several other conventional values for $L_B$ with significant
variations relative to the precision required for our comparison.
Since this parameter is the only one with an apparently significant
discrepancy between the two analyses, it seems likely that the
conversion we used was not quite correct.

\begin{figure}
	\includegraphics[width=\columnwidth]{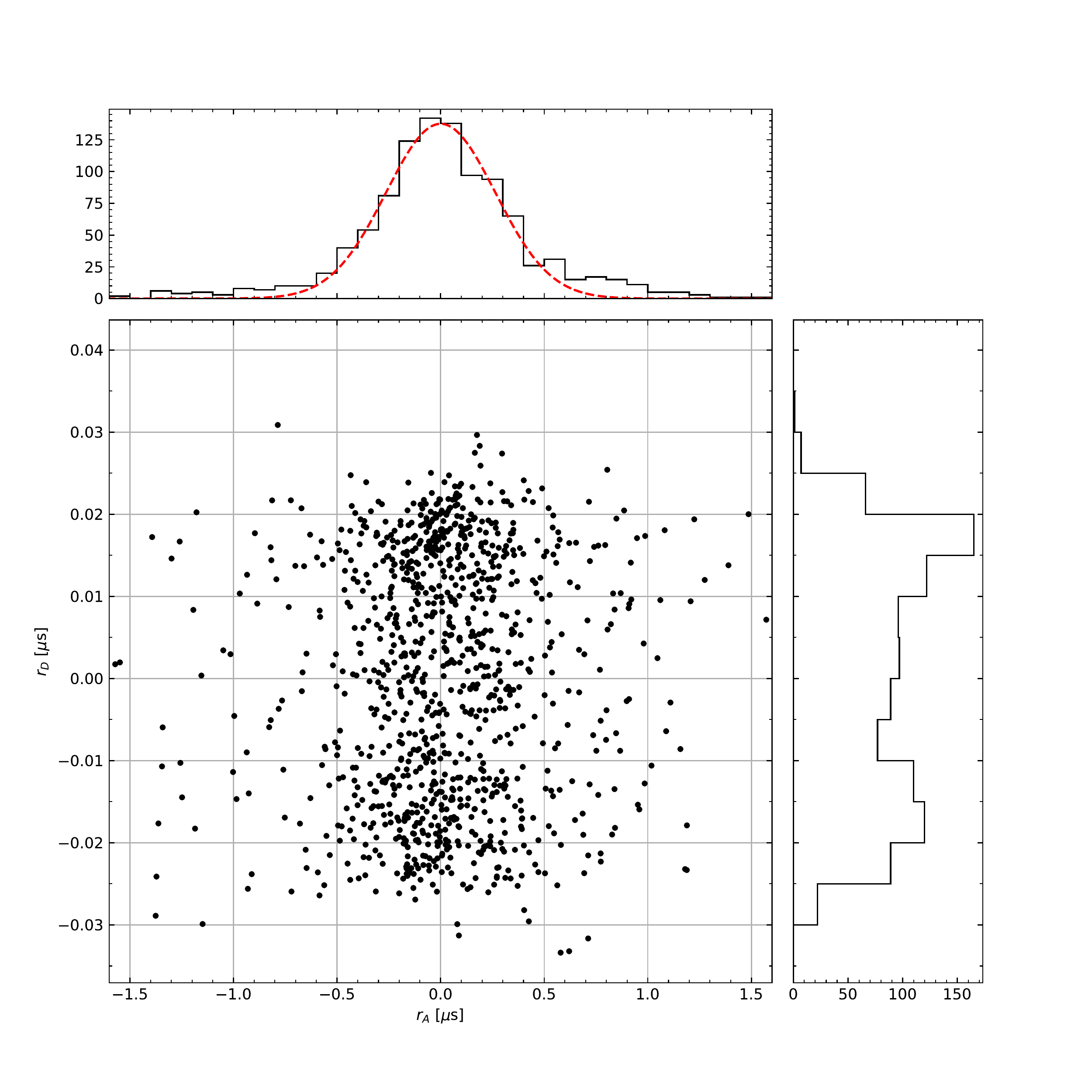}
    \caption{Scatter plot showing the averaged post-fit residuals $r_A$, and the residual differences $r_D$, along with histograms of each. A Gaussian distribution fit to the central part of the $r_A$ distribution has a SD of $0.25\,\mu$s. No significant correlation is evident between the average and difference residuals.}
    \label{fig:rArD}
\end{figure}

The post-fit residuals from the PEP and \tempo{} analyses ($r_{PEP}$ and $r_{TEMPO2}$, respectively) are very similar.  We take as the nominal, the average between these, $r_A\equiv (r_{PEP}+r_{TEMPO2})/2$.  A histogram of $r_A$ is shown in Fig.~\ref{fig:rArD}, where a Gaussian distribution is fit to the central part of the histogram yielding a standard deviation (SD) of 0.25\,$\mu$s.
By comparison, the SD of the full set of $r_A$ is 0.57\,$\mu$s.  The figure shows that the “tails” of the data distribution are over-populated, which is not unusual for real data.  Figure~\ref{fig:rArD} also shows a histogram of the difference of the two sets of residuals:  $r_D \equiv (r_{PEP}-r_{TEMPO2})$.  Because the \tempo{} analysis ties the timing model to an arbitrary arrival time of a fictitious reference pulse, whereas PEP estimates the pulse phase at a reference epoch simultaneously with the other parameters, we have adjusted the residual difference by subtracting its mean to make the two sets more directly comparable.  The nearly rectangular distribution of $r_D$ has a SD of 15\,ns, which is 38 times smaller than the SD of $r_A$.  The maximum absolute value among the $r_D$ is 33\,ns, making it unlikely that there exists a significant error in either of the two programs being compared. Finally, Fig.~\ref{fig:rArD} shows the lack of correlation between $r_A$ and $r_D$; the calculated correlation coefficient $\rho=0.02$ confirms that lack.  The plot emphasizes that the residual differences $r_D$ are small compared to the residuals $r_{PEP}$ and $r_{TEMPO2}$, which is consistent with the two software packages having nearly identical, correctly programmed models. 

\section{Conclusion} \label{sec:Conc}

We have described the Planetary Ephemeris Program (PEP) and its principal capabilities.  Over nearly six decades, this Fortran program of over $10^5$ lines has grown and evolved as scientific opportunities in the form of new data, usually of higher accuracy, and new data types have become available.  That evolution has occasionally been facilitated by the direct comparison with independently written codes with which there is a substantial overlap in capability.  We presented one such comparison (with \tempo{}) here, and mentioned another (with EPM).  PEP is still in active use and the PEP source code is publicly available via GitLab.  PEP is the only open source, general analysis tool for solar-system astrometric data in the world, as far as we are aware.

\section{Acknowledgement} \label{sec:Ack}

We thank Ryan Shannon and George Hobbs for  for help with \tempo{}. This material is based  upon work supported by the National Science Foundation under Grant No. PHY1708215, and by the National Aeronautics and Space Administration under Grant No. NNX15AC51G issued through the mission directorate, and under Grant No. NNX16AH49H through the Massachusetts Space Grant Consortium.

\appendix

This appendix describes an unusual portion of the estimation module in PEP.  It has the ability to transform (partially prereduce) the normal equations (NE) prior to obtaining a solution. The prereduced NE are smaller than the original but, for the parameters represented, yield the same parameter estimate and statistics as the full NE.  We start by discussing one motivation for having such a capability in Section A, Numerical Experiments. In Section B, Transformation of the NE, we provide a derivation of the algorithm.  \\

\section{Numerical Experiments} \label{sec:NumExp}

A dynamic model of the solar system is limited in complexity by the ability of the available data to estimated the model-related parameters without making the NE degenerate (rank deficient) or even making the condition number unreasonably large. Thus, we are always using an ``inadequate model" in our analysis, which implies that we must expect to get biased estimates. That is troubling in the case of those parameters that are of primary scientific  interest.  Many approaches have been proposed to address these concerns.  

One approach involves running a large number of solutions that differ only in the inclusion or exclusion of members of a particular subset of the model parameters.  The selection of solutions is a matter of judgement by the investigators. 

The full ensemble of parameters included in a solar-system solution often contains sets of ``nuisance parameters" that must be included but whose estimate is of little or no interest.  For example, when planetary radar observations were in regular use in the solar-system analysis, we developed an \textit{ad hoc} model of the topography of each planet in the region from which radar returns would come.

When the set of nuisance parameters becomes very large, the matrix inversion can take long enough that it interferes with the work, either because the cost or availability of machine time is limiting or because an investigator is trying to work in real time.  As computers become more powerful and people tackle more difficult problems, this problem may become or cease to be important. We developed a solution dubbed ``Partial Pre-Reduction of the Normal Equations" (PPRNE). It entails a rigorous transformation of the NE to a new set of NE such that any solution of a subset of the new NE yields the same parameter estimate variance and co-variance as would have been found for the parameters of the new NE by inverting the original NE. PPRNE is the central theme of this appendix.

\section{Transformation of the NE} \label{sec:Trans}

In classical weighted least-squares (WLS) fitting, there is a data vector $z$ of length $m$ and a correct model $H(X)$ where $X$ is a vector of parameters of length $n$. No unique solution is possible for $m<n$. We define a loss function $J=r^{\dag}R^{-1}r$, where $r=z-H(X)$ is the prefit residual, $R$ is the data noise covariance, and $^{\dag}$ indicates the transpose. When $R$ is diagonal, the diagonal elements of $R^{-1}$ become a vector of data weights. The NE of WLS fitting take the form
\begin{equation}
\label{eq:BDU}
B\Delta = U, \textup{ which has solution, } \Delta =  B^{-1} U,
\end{equation}
 where $B$ is the $n \times n$ coefficient matrix, $\Delta$ is a vector of length $n$ of adjustments, and $U$ is the vector of length $n$ that contains the error to be corrected.  In particular:
 
 \begin{eqnarray}
 B = A^{\dag}R^{-1}A\hspace{6 mm}
 U = A^{\dag}R^{-1}r\hspace{3 mm}
 \textup{ where } \hspace{3 mm} A = \frac{\partial H}{\partial X}.
 \end{eqnarray} 

 We divide the set of $n$ parameters into two sets, $\alpha$ and $\beta$ such that all of the ``interesting parameters" are in the set $\alpha$. This set includes all parameters whose estimate and uncertainties are of prime interest and all additional parameters that one may want to sometimes include in the analysis and other times may want to exclude from the analysis. The set $\beta$ includes all nuisance parameters and all additional parameters that need to be part of the solution but can be ignored in the planned analysis.
 
 We can order the parameters such that all $\alpha$ parameters precede all $\beta$ parameters. Then the vectors and matrices of Eq.~(\ref{eq:BDU}) can be written in partitioned form:
 
 \begin{eqnarray}
 B = \left(\begin{array}{cc}
 C & F\\
 F^{\dag} & D
 \end{array} \right),\hspace{4 mm}
 B^{-1}=\left(\begin{array}{cc}
 P & Q\\
 Q^{\dag} & R
 \end{array} \right), \hspace{4 mm}
 \Delta = \left(\begin{array}{cc}
 Y\\ Z
 \end{array} \right),\hspace{4 mm}
 U = \left(\begin{array}{cc}
 V\\ W
 \end{array} \right).  
 \end{eqnarray} 
 From this it follows that
 
 \begin{equation}
 V = C Y + F Z , \hspace{4 mm} 
 W = F^{\dag} Y + D Z, \hspace{4 mm} 
 Z = D^{-1}(W-F^{\dag} Y).  \\
 \end{equation} 
By combining these we obtain

\begin{equation}
(C-FD^{-1}F^{\dag})Y=V-FD^{-1}W,
\end{equation}
and thus
\begin{equation}
\label{eq:YCV}
Y=\bar{C}^{-1}\bar{V},
\end{equation}
where

\begin{equation}
 \bar{C}=C-FD^{-1}F^{\dag}, \hspace{4 mm} 
 \bar{V}=V-FD^{-1}W.
\end{equation}
Equation~(\ref{eq:YCV}) shows that the full NE can be replaced by a smaller set that yields the same adjustment for the parameters in the $\alpha$ set as would be obtained from the inversion of the full NE. We next show that the variance and co-variance are also given correctly by the smaller NE.

\begin{equation}
 B^{-1}B=\left(\begin{array}{cc}
 PC+QF^{\dag} & PF+QD\\
 Q^{\dag}C+RF^{\dag} & Q^{\dag}F+RD
 \end{array} \right)=
 \left(\begin{array}{cc}
 I_{\alpha} & 0\\
 0 & I_{\beta}
 \end{array} \right)
\end{equation}
To show that the PPRNE yields the same statistics as the original NE, we need to show that $\bar{C}^{-1}=P$. Then:
\begin{equation}
P\bar{C}=P(C-FD^{-1}F^{\dag})=PC-PFD^{-1}F^{\dag}=PC+QDD^{-1}F^{\dag}=PC+QF^{\dag}=I_{\alpha}  
\end{equation}

\section{Change of the Weighted Sum Squared Residuals} \label{sec:WSSR}
A metric in parameter space that yields a measure of the size $N$ of the adjustment in the case of linear WLS fitting is the coefficient matrix $B$.
\begin{equation}
\label{eq:Nsquared}
N^{2}=\Delta^{\dag}B\Delta=\Delta^{\dag}U
\end{equation}
For the predicted residual $r_p$, which is the post-fit residual in the linear case, 
\begin{equation}
r_p = r-A\Delta = r-A(A^{\dag}R^{-1}A)^{-1} A^{\dag}R^{-1}r = (I-G)r
\end{equation}
where $G = A(A^{\dag}R^{-1}A)^{-1} A^{\dag}R^{-1}$ is a projection operator, $G = GG$. Then the sum squared weighted post-fit residual is
\begin{equation}
r_p^{\dag} R^{-1} r_p = r^{\dag}(I-G^{\dag})R^{-1}(I-G)r.
\end{equation}
We note that
\begin{equation}
G^{\dag}R^{-1}=R^{-1}A(A^{\dag}R^{-1}A)^{-1} A^{\dag}R^{-1} = R^{-1}G
\end{equation}
and therefore
\begin{equation}
r_p^{\dag} R^{-1} r_p = r^{\dag}R^{-1}(I-G)(I-G)r=r^{\dag}R^{-1}r - r^{\dag}R^{-1}Gr.
\end{equation}
Revisiting Eq.~(\ref{eq:Nsquared}) we find
\begin{equation}
N^{2}=\Delta^{\dag}B\Delta=r^{\dag}R^{-1}Gr
\end{equation}
\begin{equation}
r^{\dag}R^{-1}r - r_p^{\dag} R^{-1} r_p =  N^{2}.
\end{equation}
Thus, the effect of a fit is that the sum-squared of the residuals drop (between pre-fit and post-fit) by $\Delta^{\dag}U$, which is easily calculated without first calculating the predicted residuals. 

We next apply this analysis of residuals to the PPRNE. With the partitioning of the matrices and vectors into $\alpha$ and $\beta$ parts, Eq.~(\ref{eq:Nsquared}) becomes,
\begin{equation}
N^{2}=\Delta^{\dag}U=Y^{\dag}\bar{V}+N_{\beta}^2, \hspace{4 mm} 
N_{\beta}^2=W^{\dag}D^{-1}W.
\end{equation}
The quantity $N_{\beta}^2$ would be calculated once, say, at the time the NE were Partially Pre-Reduced.\\

\bibliography{PEP}{}
\bibliographystyle{aasjournal}

\end{document}